\documentclass[preprint,preprintnumbers,amsmath,amssymb]{revtex4}
\usepackage{graphicx}
\usepackage{color}

\def\beq{\begin{equation}}
\def\eeq{\end{equation}}
\def\eeqn{\end{equation}}
\newcommand\iden{\leavevmode\hbox{\small1\normalsize\kern-.33em1}}

\newcommand{\bea} {\begin{eqnarray}}
\newcommand{\eea} {\end{eqnarray}}

\let\jnfont=\rm
\def\NPB#1,{{\jnfont Nucl.\ Phys.\ B }{\bf #1},}
\def\PLB#1,{{\jnfont Phys.\ Lett.\ B }{\bf #1},}
\def\EPJC#1,{{\jnfont Eur.\ Phys.\ Jour.\ C }{\bf #1},}
\def\PRD#1,{{\jnfont Phys.\ Rev.\ D }{\bf #1},}
\def\PRL#1,{{\jnfont Phys.\ Rev.\ Lett.\ }{\bf #1},}
\def\MPLA#1,{{\jnfont Mod.\ Phys.\ Lett.\ A }{\bf #1},}
\def\JPG#1,{{\jnfont J.\ Phys.\ G }{\bf #1},}
\def\CTP#1,{{\jnfont Commun.\ Theor.\ Phys.\ }{\bf #1},}
\def\JHEP#1,{{\jnfont JHEP \ }{\bf #1},}
\def\NPPS#1,{{\jnfont Nucl.\ Phys.\ Proc.\ Suppl.\ }{\bf #1},}
\def\CPC#1,{{\jnfont Computl.\ Phys.\ Commun.\ }{\bf #1},}
\def\CPL#1,{{\jnfont Chin.\ Phys.\ Lett. }{\bf #1},}
\def\AJS#1,{{\jnfont Astrophys.\ J.\ Suppl. }{\bf #1},}
\def\PR#1,{{\jnfont Phys.\ Rept. }{\bf #1},}
\def\AP#1,{{\jnfont Astropart.\ Phys. }{\bf #1},}
\def\EPL#1,{{\jnfont Europhys.\ Lett. }{\bf #1},}
\def\FP#1,{{\jnfont Fortsch.\ Phys. }{\bf #1},}
\def\chinpc#1,{{\jnfont Chin.\ Phys.\ C }{\bf #1},}

\begin{document}

\title{Lepton-specific inert two-Higgs-doublet model confronted with the new results for muon and electron g-2 anomalies and multi-lepton searches at the LHC}
\renewcommand{\thefootnote}{\fnsymbol{footnote}}

\author{Xiao-Fang Han$^{1}$, Tianjun Li$^{2,3}$, Hong-Xin Wang$^{1}$, Lei Wang$^{1}$, Yang Zhang$^{4}$}
 \affiliation{$^1$ Department of Physics, Yantai University, Yantai
264005, P. R. China\\
$^2$ CAS Key Laboratory of Theoretical Physics, Institute of Theoretical Physics,
 Chinese Academy of Sciences, Beijing 100190,  P. R. China\\
$^3$ School of Physical Sciences, University of Chinese Academy of Sciences, Beijing 100049,  P. R.  China\\
$^4$ School of Physics and Microelectronics, Zhengzhou University, ZhengZhou 450001, P. R. China}
\renewcommand{\thefootnote}{\arabic{footnote}}

\date{\today}

\begin{abstract}
Combining the multi-lepton searches at the LHC, we study the possibilities of accommodating
the new data of muon and electron $g-2$ anomalies in the lepton-specific inert two-Higgs-doublet model.
We take the heavy CP-even Higgs as the 125 GeV Higgs, and find the muon and electron $g-2$ anomalies can be explained simultaneously 
in the region of 5 GeV $< m_h<60$ GeV, 200 GeV $<m_A< 620$ GeV, 190 GeV $<m_{H^\pm}< 620$ GeV for appropriate Yukawa couplings between
 leptons and inert Higgs. Meanwhile, the model can give a better fit to the data of lepton universality in $\tau$ decays than the SM.
Further, the multi-lepton event searches at the LHC impose a stringent upper bound on $m_h$, $m_h<$ 35 GeV.
\end{abstract}
\maketitle

\section{Introduction}
The Fermilab collobartion presented their new result for
muon anomalous magnetic moment which now, combined with the
data of the E821 \cite{mug2-exp-1,mug2-exp-2}, amounts to \cite{fermig2}
\bea
\Delta a_\mu=a_\mu^{exp}-a_\mu^{SM}=(251\pm59)\times10^{-11}.
\eea 
The experimental value has an
approximate $4.2\sigma$ discrepancy from the SM prediction \cite{g2qed,g2ew,g2light,g2hvp}.
Besides, an improvement in the measured mass of atomic Cesium
used in conjunction with other known mass ratios and
the Rydberg constant leads to the most precise value of the fine structure
constant \cite{alpha-exp}. As a result, the experimental value of the electron $g-2$ has a
$2.4\sigma$ deviation from the SM prediction \cite{eg2-2.4-1,eg2-2.4-2}, 
\bea
\Delta a_e=a_e^{exp}-a_e^{SM}=(-87\pm36)\times10^{-14},
\eea
which is opposite in sign from the muon $g-2$.

The lepton-specific two-Higgs-doublet model (2HDM) can explain muon $g-2$ anomaly simply \cite{mu2h1,mu2h2,mu2h3,mu2h5,mu2h9,mu2h11,mu2h12,mu2h13,mu2h14,
mu2h15,tavv-1,crivellin,tavv-2,mu2h19,mu2h20,mu2h21}, 
 but will raise the discrepancy in lepton flavor universality (LFU) in $\tau$ decay \cite{tavv-1,crivellin,tavv-2}.
 In addition, the muon and electron $g-2$ anomalies can not be explained simultaneously in lepton-specific 2HDM since there is an opposite sign between them.
There have been some new physics models which attempt to explain the muon and electron $g-2$ 
simultaneously \cite{mueg1,mueg2,mueg3,mueg4,mueg5,mueg6,mueg7,mueg8,mueg9,mueg10,mueg11,mueg12,mueg13,mueg14,mueg15,mueg16,
mueg17,mueg18,mueg19,mueg20,mueg21,mueg22,mueg23,mueg24,mueg25,mueg26,mueg27,mueg28}. In Ref. \cite{mueg5}, we propose a lepton-specific inert 2HDM 
to explain the muon and electron $g-2$ anomalies, and the key point is that these Yukawa couplings for $\mu$ and $e/\tau$
have opposite sign. In the model, the extra Higgses will decay into leptons mainly. Therefore, the 
multi-lepton event searches at the LHC can impose stringent constraints on the model. Motivated by the new results for muon $g-2$ and 
multi-lepton event searches at the LHC, we revisit the possibilities of accommodating
 muon and electron $g-2$ anomalies in the lepton-specific inert 2HDM. In this paper, we study a different scenario from that of Ref. \cite{mueg5}, 
and take the inert CP-even Higgs $h$ to be lighter than the SM-like Higgs $H$. In such scenario, a very light CP-even Higgs $h$ plays important roles in explaining the
muon and electron g-2 anomalies, and is expected to more easily avoid the constraints from the multi-lepton searches at the LHC than
the scenario of Ref. \cite{mueg5} in which $h$ is heavier than the 125 GeV Higgs. 

Our work is organized as follows. In Sec. II we will give a brief introduction on the model. 
 In Sec. III and Sec. IV, we introduce the numerical calculations, and show the allowed and excluded parameter space.
Finally, we give our conclusion in Sec. V.

\section{Lepton-specific inert 2HDM} 
We add an inert Higgs doublet field to the SM, and the scalar potential is written as,
\begin{eqnarray} \label{V2HDM} \mathrm{V} &=& Y_1
(\Phi_1^{\dagger} \Phi_1) + Y_2 (\Phi_2^{\dagger}
\Phi_2)+ \frac{\lambda_1}{2}  (\Phi_1^{\dagger} \Phi_1)^2 +
\frac{\lambda_2}{2} (\Phi_2^{\dagger} \Phi_2)^2  \nonumber \\
&&+ \lambda_3
(\Phi_1^{\dagger} \Phi_1)(\Phi_2^{\dagger} \Phi_2) + \lambda_4
(\Phi_1^{\dagger}
\Phi_2)(\Phi_2^{\dagger} \Phi_1)\nonumber \\
&&+ \left[\frac{\lambda_5}{2} (\Phi_1^{\dagger} \Phi_2)^2 + \rm
h.c.\right].
\end{eqnarray}
Here we impose a discrete $Z_2$ symmetry under which $\Phi_2$ is odd and the SM fields are even.
We study the CP-conserving case where all
$\lambda_i$ are real. The two complex
scalar doublets with the hypercharge $Y = 1$ can be given as
\begin{equation} \label{field}
\Phi_1=\left(\begin{array}{c} G^+ \\
\frac{1}{\sqrt{2}}\,(v+H+iG_0)
\end{array}\right)\,, \ \ \
\Phi_2=\left(\begin{array}{c} H^+ \\
\frac{1}{\sqrt{2}}\,(h+iA)
\end{array}\right).
\end{equation}
The $\Phi_1$ field has the vacuum expectation value (VEV) $v=$246
GeV, and the VEV of $\Phi_2$ field is zero. The $Y_1$ is determined by the scalar
potential minimum condition,
\beq
Y_1=-\frac{1}{2}\lambda_1 v^2.
\eeq

The Nambu-Goldstone bosons $G^0$ and $G^+$ are eaten by the gauge bosons. The
$H^+$ and $A$ are the mass eigenstates of the charged Higgs boson and
CP-odd Higgs boson.  
 Their masses are given as
\beq \label{masshp}
 m_{H^\pm}^2  = Y_2+\frac{\lambda_3}{2} v^2, ~~m_{A}^2  = m_{H^\pm}^2+\frac{1}{2}(\lambda_4-\lambda_5) v^2.
 \eeq
The $H$ is the SM-like Higgs, and has no mixing with the inert CP-even Higgs $h$. Their masses are given as
\beq \label{massh}
 m_{H}^2  = \lambda_1 v^2\equiv (125~{\rm GeV })^2, ~~m_{h}^2  = m_{A}^2+\lambda_5 v^2.
 \eeq

The fermions obtain the mass terms from the Yukawa interactions with $\Phi_1$,
 \beq \label{yukawacoupling} - {\cal L} = y_u\overline{Q}_L \,
\tilde{{ \Phi}}_1 \,u_R + y_d\overline{Q}_L\,{\Phi}_1 \, d_R +  y_l\overline{L}_L \, {\Phi}_1
\, e_R + \mbox{h.c.}, \eeq
where $Q_L^T=(u_L\,,d_L)$, $L_L^T=(\nu_L\,,l_L)$, 
$\widetilde\Phi_{1}=i\tau_2 \Phi_{1}^*$, and $y_u$, $y_d$ and $y_\ell$ are
$3 \times 3$ matrices in family space. In addition, in the lepton sector we introduce 
the $Z_2$ symmetry-breaking Yukawa interactions of $\Phi_2$,
 \bea \label{phi2coupling} - {\cal L} &=& \frac{\sqrt{2}m_e}{v}\, \kappa_e \,\overline{L}_{1L} \, {\Phi}_2
\, e_R  \, +\frac{\sqrt{2}m_\mu}{v}\, \kappa_\mu\, \overline{L}_{2L} \, {\Phi}_2
\,\mu_R \, \nonumber\\
&&+\frac{\sqrt{2}m_\tau}{v}\, \kappa_\tau \,\overline{L}_{3L} \, {\Phi}_2
\,\tau_R\, + \, \mbox{h.c.}\,. \eea
We can obtain the lepton Yukawa couplings of extra Higgses ($h$, $A$, and $H^\pm$) 
from Eq. (\ref{phi2coupling}).
The neutral Higgses $h$ and $A$ have no couplings to $ZZ, ~WW$.
At the tree-level, the SM-like Higgs $H$ has the same couplings to fermions and gauge bosons as the SM.

\section{Numerical calculations}
In the lepton-specific 2HDM and aligned 2HDM, $\kappa_\tau$ equals to $\kappa_\mu$. As a result, the decay $\tau\to \mu\nu\nu$
will obtain negative contribution from the diagram mediated by the charged Higgs.
In the model we take $\kappa_\mu$ to be opposite in sign from $\kappa_\tau$. Thus, the diagrams mediated
by the charged Higgs produce positive contribution to the decay $\tau\to \mu\nu\nu$, and the model
can give better fit to the data of the LUF in the $\tau$ decay.
When $\kappa_\mu$ is opposite in sign from $\kappa_\tau$, the contributions of the CP-even (CP-odd) Higgses to muon $g-2$
are positive (negative) at the two-loop level and positive (negative) at one-loop level.
Therefore, we take the heavy CP-even Higgs as the 125 GeV Higgs, $m_H=125$ GeV, and make the light CP-even Higgs $h$ to be light enough
to enhance muon $g-2$. Since electron $g-2$ is opposite in sign from muon $g-2$, we will take $\kappa_e$ to have a opposite sign
from $\kappa_\mu$. 

In our calculations, we take $\lambda_2$, $\lambda_3$, $m_H$, $m_h$, $m_A$ and $m_{H^\pm}$ as the input parameters.
According to Eqs. (\ref{masshp}, \ref{massh}), the values of $\lambda_1$, $\lambda_5$ and $\lambda_4$ can be 
determined. $\lambda_2$ controls the quartic couplings of extra Higgses, and does not affect the observables considered in our paper.
We take $\lambda_3=\lambda_4+\lambda_5$ to make the $Hhh$ coupling to be absent.

We scan over several key parameters in the following ranges:
\bea
&&5{\rm GeV}<m_h<60~{\rm GeV}, ~200{\rm GeV} <m_A<800{\rm GeV},~ 90{\rm GeV} <m_{H^{\pm}}<800{\rm GeV},\nonumber\\
&&1 <\kappa_\tau<140,~-200<\kappa_\mu<-1,~1<\kappa_e<500.
\eea
In such ranges of $\kappa_\tau$, $\kappa_\mu$ and $\kappa_e$, the corresponding Yukawa couplings do not become non-perturbative.

The model gives the new contributions to muon $g-2$ via the
one-loop diagrams and the two-loop Barr-Zee diagrams involving extra Higgses. For the
one-loop contributions \cite{mu2h1} we have
 \beq
    \Delta a_\mu^{\mbox{$\scriptscriptstyle{\rm 2HDM}$}}({\rm 1loop}) =
    \frac{m_\mu^2}{8 \pi^2 v^2} \, \sum_i
     \kappa_\mu^2 \, r_{\mu}^i \, F_j(r_{\mu}^i),
\label{amuoneloop}
\end{equation}
where $i = h,~ A ,~ H^\pm$, $r_{\mu}^ i =  m_\mu^2/M_j^2$. For
$r_{\mu}^i\ll$ 1 we have
\beq
    F_{h}(r) \simeq- \ln r - 7/6,~~
    F_A (r) \simeq \ln r +11/6, ~~
    F_{H^\pm} (r) \simeq -1/6.
    \label{oneloopintegralsapprox3}
\eeq
The contributions of the two-loop diagrams with a closed fermion loop are given by  
\beq
    \Delta a_\mu^{\mbox{$\scriptscriptstyle{\rm 2HDM}$}}({\rm 2loop})
    = \frac{m_{\mu}^2}{8 \pi^2 v^2} \, \frac{\alpha_{\rm em}}{\pi}
    \, \sum_{i,\ell} \, Q_\ell^2  \,  \kappa_\mu  \, \kappa_\ell \,  r_{\ell}^i \,  G_i(r_{\ell}^i),
\label{barr-zee}
\end{equation}
where $i = h,~ A$, $\ell=\tau$, and $m_\ell$ and $Q_\ell$ are the mass and
electric charge of the lepton $\ell$ in the loop. The functions $G_i(r)$ are \cite{mu2h2,mu2h3}
\begin{eqnarray}
    && G_{h}(r) = \int_0^1 \! dx \, \frac{2x (1-x)-1}{x(1-x)-r} \ln
    \frac{x(1-x)}{r}, \\
    && G_{A}(r) = \int_0^1 \! dx \, \frac{1}{x(1-x)-r} \ln
    \frac{x(1-x)}{r}.
\end{eqnarray}
In our calculation, we also include the contributions of the two-loop diagrams with a closed charged Higgs loop,
and find that their contributions are much smaller than those of fermion loop. 
The calculation of $\Delta a_e$ is similar to that of $\Delta a_\mu$, and we include the contributions of the two-loop diagrams 
with closed $\mu$ loop and $\tau$ loop.

The HFAG collaboration reported three ratios from pure leptonic processes, and two ratios
from semi-hadronic processes, $\tau \to \pi/K \nu$ and $\pi/K \to \mu \nu$ \cite{tauexp}:
\begin{eqnarray} \label{hfag-data}
&&
\left( g_\tau \over g_\mu \right) =1.0011 \pm 0.0015,~\left( g_\tau \over g_e \right)  = 1.0029 \pm 0.0015,\nonumber\\
&&\left( g_\mu \over g_e \right) = 1.0018 \pm 0.0014,~\left( g_\tau \over g_\mu \right)_\pi = 0.9963 \pm 0.0027,\nonumber\\
&&\left( g_\tau \over g_\mu \right)_K = 0.9858 \pm 0.0071,
\end{eqnarray}
with the following definitions
\begin{eqnarray} 
&&
\left( g_\tau \over g_\mu \right)^2 \equiv \bar{\Gamma}(\tau\to e
\nu\bar{\nu})/\bar{\Gamma}(\mu\to e \nu\bar{\nu}),\nonumber\\
&&\left( g_\tau \over g_e \right)^2  \equiv \bar{\Gamma}(\tau\to \mu
\nu\bar{\nu})/\bar{\Gamma}(\mu\to e \nu\bar{\nu}),\nonumber\\
&&\left( g_\mu \over g_e \right)^2 \equiv \bar{\Gamma}(\tau\to \mu
\nu\bar{\nu})/\bar{\Gamma}(\tau\to e \nu\bar{\nu}).
\end{eqnarray}
Where $\bar{\Gamma}$ denotes the partial width normalized to its SM value.
The correlation matrix for the above five observables is 
\begin{equation} \label{hfag-corr}
\left(
\begin{array}{ccccc}
1 & +0.53 & -0.49 & +0.24 & +0.12 \\
+0.53  & 1     &  + 0.48 & +0.26    & +0.10 \\
-0.49  & +0.48  & 1       &   +0.02 & -0.02 \\
+0.24  & +0.26  & +0.02  &     1    &     +0.05 \\
+0.12  & +0.10  & -0.02  &  +0.05  &   1 
\end{array} \right).
\end{equation}

In the model, we have the ratios
\begin{eqnarray} \label{deltas-data}
&&\left( g_\tau \over g_\mu \right)^2\approx \frac{1+ 2\delta^\tau_{\rm loop}}{1+ 2\delta^\mu_{\rm loop}}, \nonumber\\
&&\left( g_\tau \over g_e \right)^2   \approx \frac{1+ 2\delta_{\rm tree}+ 2\delta^\tau_{\rm loop}}{1+2\delta^\mu_{\rm loop}}, \nonumber\\
&&\left( g_\mu \over g_e \right)^2   \approx 1+ 2\delta_{\rm tree}, 
\nonumber\\
&&
\left( g_\tau \over g_\mu \right)_\pi^2= \left( g_\tau \over g_\mu \right)_K^2=\left( g_\tau \over g_\mu \right)^2.
\end{eqnarray} 
The $\delta_{\rm tree}$ and $\delta_{\rm loop}^{\tau,\mu}$ are respectively corrections from
the tree-level diagrams and the one-loop diagrams mediated by the charged Higgs. They are given as \cite{tavv-1,tavv-2,mu2h19}
\begin{eqnarray} \label{tree-tau}
\delta_{\rm tree} &=& {m_\tau^2 m_\mu^2 \over 8 m^4_{H^\pm}} \kappa^2_\tau \kappa^2_\mu
- {m_\mu^2 \over m^2_{H^\pm}} \kappa_\tau \kappa_\mu {g(m_\mu^2/m^2_\tau) \over f(m_\mu^2/m_\tau^2)}, \\
\delta_{\rm loop}^{\tau,\mu} &=& {1 \over 16 \pi^2} { m_{\tau,\mu}^2 \over v^2}  \kappa^2_{\tau,\mu}
\left[1 + {1\over4} \left( H(x_A)  + H(x_h)\right)
\right]\,, 
\end{eqnarray}
where $f(x)\equiv 1-8x+8x^3-x^4-12x^2 \ln(x)$, $g(x)\equiv 1+9x-9x^2-x^3+6x(1+x)\ln(x)$ and
$H(x_\phi) \equiv \ln(x_\phi) (1+x_\phi)/(1-x_\phi)$ with $x_\phi=m_\phi^2/m_{H^{\pm}}^2$.

We perform $\chi^2_\tau$ calculation for the five observables. The covariance matrix constructed from the data of Eq. (\ref{hfag-data})
and Eq. (\ref{hfag-corr}) has a vanishing eigenvalue, and the corresponding degree is removed in our calculation.
In our discussions we require the value of $\chi^2_\tau$ to be smaller than the SM value, $\chi^2_\tau({\rm SM})=12.3$.

The measured values of the ratios of the leptonic $Z$ decay
branching fractions are given as \cite{zexp}:
\begin{eqnarray}
{\Gamma_{Z\to \mu^+ \mu^-}\over \Gamma_{Z\to e^+ e^- }} &=& 1.0009 \pm 0.0028
\,,\nonumber\\ 
{\Gamma_{Z\to \tau^+ \tau^- }\over \Gamma_{Z\to e^+ e^- }} &=& 1.0019 \pm 0.0032
\,, \label{lu-zdecay}
\end{eqnarray}
with a correlation of $+0.63$. 
In the model, the width of $Z\to \tau^+\tau^-$ can have sizable deviation from the SM value
due to the loop contributions of the extra Higgs bosons, because they strongly interact with charged
leptons. The quantities of Eq. (\ref{lu-zdecay}) are calculated in the model are similar to Refs. \cite{mu2h19,tavv-1,tavv-2}.
\begin{eqnarray} 
&&{\Gamma_{Z\to \mu^+ \mu^-}\over \Gamma_{Z\to e^+ e^- }} \approx 1.0+ {2 g_L^e{\rm Re}(\delta g^{\rm 2HDM}_L)+ 2 g_R^e{\rm Re}(\delta g^{\rm 2HDM}_R) \over {g_L^e}^2 + {g_R^e}^2 }\frac{m_\mu^2\kappa_\mu^2}{m_\tau^2\kappa_\tau^2},\nonumber\\ 
&&{\Gamma_{Z\to \tau^+ \tau^- }\over \Gamma_{Z\to e^+ e^- }} 
\approx 1.0+ {2 g_L^e{\rm Re}(\delta g^{\rm 2HDM}_L)+ 2 g_R^e{\rm Re}(\delta g^{\rm 2HDM}_R) \over {g_L^e}^2 + {g_R^e}^2 }\,.
\,
\end{eqnarray}
where the SM value $g_L^e=-0.27$ and $g_R^e=0.23$. $\delta g^{\rm 2HDM}_L$ and $\delta g^{\rm 2HDM}_R$ are given as 
\begin{eqnarray} \label{dgLR}
\delta g^{\rm 2HDM}_L &=& {1\over 16\pi^2} {m_\tau^2 \over v^2} \kappa_\tau^2 \,
\bigg\{
 -{1\over2} B_Z(r_A)- {1\over2} B_Z(r_h) -2 C_Z(r_A, r_h)
   \nonumber \\
&&  + s_W^2 \left[ B_Z(r_A) + B_Z(r_h) + \tilde C_Z(r_A) + \tilde C_Z(r_h) \right] \bigg\} 
\,,\nonumber\\ 
\delta g^{\rm 2HDM}_R &=& {1\over 16\pi^2} {m_\tau^2 \over v^2} \kappa_\tau^2 \,
\bigg\{ 2 C_Z(r_A, r_h) - 2 C_Z(r_{H^\pm}, r_{H^\pm}) 
+ \tilde C_Z(r_{H^\pm}) 
\nonumber \\
&& - {1\over2} \tilde C_Z(r_A)  - {1\over2} \tilde C_Z(r_h)      
 +  s_W^2 \left[ B_Z(r_A) + B_Z(r_h) + 2 B_Z(r_{H^\pm})\right.
\nonumber \\
&& \left. +\tilde C_Z(r_A) + \tilde C_Z(r_h) + 4C_Z(r_{H^\pm},r_{H^\pm}) \right] \bigg\} 
\,,
\end{eqnarray}
where $r_\phi = m_\phi^2/m_Z^2$ with $\phi=A,h, H^\pm$, and 
\begin{eqnarray} 
\label{loopftn}
B_Z(r) &=& -{\Delta_\epsilon \over 2} -{1\over4} + {1\over2} \log(r) \,,\\ 
C_Z(r_1,r_2) &=& {\Delta_\epsilon \over4} -{1\over2} \int^1_0 d x \int^x_0 d y\,
\log[ r_2 (1-x) + (r_1 -1) y + x y] \,,\\ 
\tilde C_Z(r) &=& {\Delta_\epsilon \over2}+{1\over2}  - r\big[1+\log(r) \big]
+r^2 \big[ \log(r) \log(1+r^{-1}) \nonumber\\
&&
-{\rm Li_2}(-r^{-1}) \big] -{i \pi\over2}
\left[ 
	1 - 2r + 2r^{2}\log(1+r^{-1})
\right].
\end{eqnarray}

The 125 GeV Higgs $(H)$ has the same tree-level couplings to the fermions and gauge bosons as the SM, and the $H\to hh$ decay
is absent for $\lambda_3=\lambda_4+\lambda_5$.
Since the extra Higgses have no coupling to quarks, we can safely neglect the constraints from the meson observable.
 We employ the $\textsf{2HDMC}$ \cite{2hc-1}
to implement theoretical constraints from the vacuum stability, unitarity and
coupling-constant perturbativity, as well as the constraints from
the oblique parameters ($S$, $T$, $U$). Adopting the recent fit results in Ref. \cite{pdg2018}, we use the following 
values of $S$, $T$, $U$,
\beq
S=0.02\pm 0.10, ~~T=0.07\pm 0.12,~~U=0.00 \pm 0.09. 
\eeq
The correlation coefficients are given by
\beq
\rho_{ST} = 0.92, ~~\rho_{SU} = -0.66, ~~\rho_{TU} = -0.86.
\eeq
$\textsf{HiggsBounds}$ \cite{hb1} is employed to implement the exclusion
constraints from the searches for the neutral and charged Higgs at the LEP 
at 95\% confidence level.

The extra Higgs bosons are dominantly produced at the LHC via
the following electroweak processes:
\bea
&&pp\to  Z^* \to hA, \label{process2}\\
&&pp\to  W^{\pm *} \to H^\pm h, \label{process3}\\
&&pp\to  W^{\pm *} \to H^\pm A, \label{process1}\\
&&pp\to  Z^*/\gamma^* \to H^+H^-, \label{process4}\\
&&pp\to  Z \to \tau^+ \tau^- h.
\eea
In our scenario, the main decay modes of the Higgs bosons are 
\bea
&&h\to \tau^+\tau^-,~\mu^+\mu^-,\cdots\cdots,\\
&&A\to  \tau^+\tau^-,~Zh,~\mu^+\mu^-\cdots\cdots,\\
&&H^\pm\to  \tau^\pm\nu,~W^\pm h,~ \mu^\pm\nu,\cdots\cdots.
\eea
We perform simulations for the processes using \texttt{MG5\_aMC-2.4.3}~\cite{Alwall:2014hca} 
with \texttt{PYTHIA6}~\cite{Torrielli:2010aw} and 
\texttt{Delphes-3.2.0}~\cite{deFavereau:2013fsa}, and impose the constraints 
from all the analysis at the 13 TeV LHC in the latest \texttt{CheckMATE 2.0.28}~\cite{Dercks:2016npn}, as well as analysis we implemented in our previous works~\cite{Pozzo:2018anw,mueg5}.
Besides, we also impose the recently published analyses of searching for events with three or more leptons, with up to two hadronical $\tau$ leptons, using 13~TeV LHC 137~fb$^{-1}$ data~\cite{CMS:2021bra}. It improves the limits on chargino mass in a  simplified SUSY model that wino-like chargino/neutralino decaying to $\tau$s. The signal regions of \texttt{4lI}, \texttt{4lJ} and \texttt{4lK} give strongest constraints on our samples, which require 4 leptons with one or two hadronical $\tau$ leptons in the final states, because of the dominated multi-lepton final states in our model.

\section{Results and discussions}
\begin{figure*}[tb]
%\begin{center}
\includegraphics[width=16.cm]{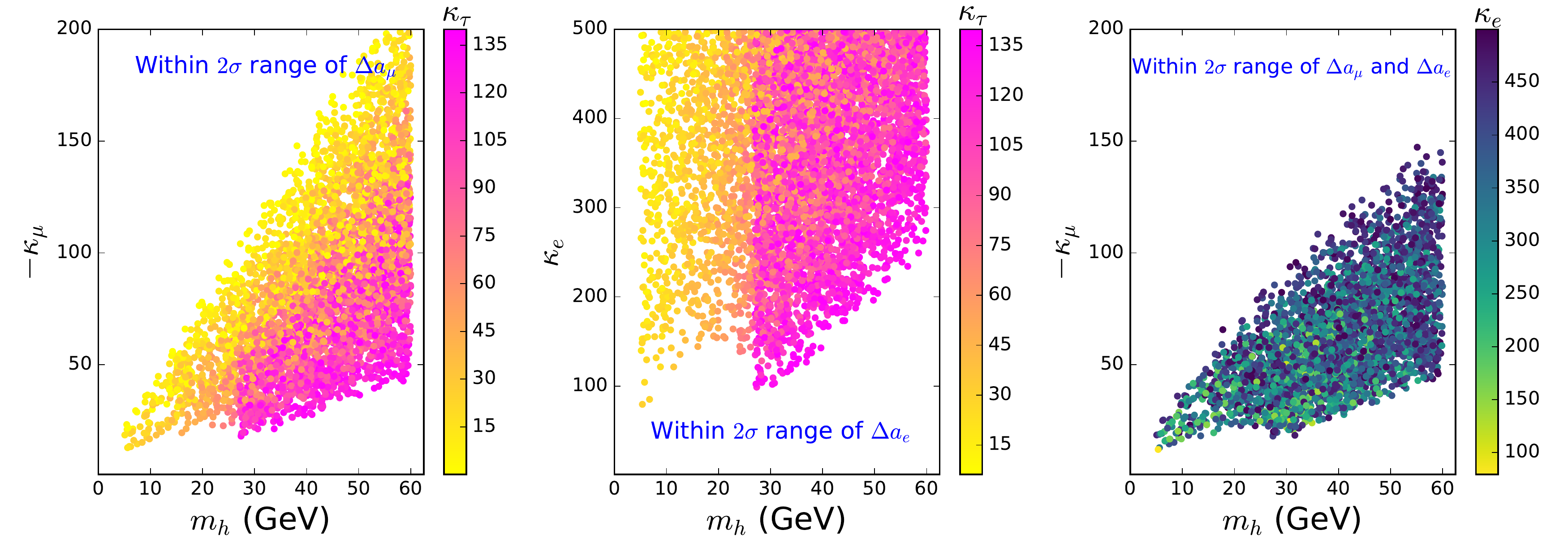}
%\end{center}
\vspace{-0.4cm}\caption{The samples within $2\sigma$ ranges of $\Delta a_\mu$ (left panel), $\Delta a_e$ (middle panel), and both
$\Delta a_\mu$ and $\Delta a_e$ (right panel). All the samples satisfy the constraints of
"pre-muon $g-2$".}
\label{meg2}
\end{figure*}

After imposing the constraints of "pre-muon $g-2$" (denoting the theory, 
the oblique parameters, the exclusion limits from the searches for Higgs at LEP), in Fig. \ref{meg2} we show the surviving samples which are 
consistent with $\Delta a_\mu$ and $\Delta a_e$ at the $2\sigma$ level.
Both one-loop and two-loop diagrams  
give positive contributions to $\Delta a_\mu$.
For $\Delta a_e$, the contributions of one-loop are positive and those of two-loop are negative.
Only the contributions of two-loop can make $\Delta a_e$ to be within $2\sigma$ range for large enough $\kappa_\tau \kappa_e$.
$\Delta a_\mu$ and $\Delta a_e$ respectively favor
$-\kappa_\mu$ and $\kappa_e$ to increase with $m_h$. For $\kappa_\tau=140$ and $m_h=60$ GeV,
$-\kappa_\mu$ and $\kappa_e$ are respectively required to be larger than 40 and 250. Due to the constraints from the
searches for $ee\to \tau\tau (h)\to \tau\tau\tau\tau$ at LEP \cite{0410017}, $\kappa_\tau$ is required to be smaller than 90
for $m_h<27$ GeV. As a result, the relative large $-\kappa_\mu$ and $\kappa_e$ are respectively required to explain 
$\Delta a_\mu$ and $\Delta a_e$ for $m_h<27$ GeV. The right panel shows that the upper limits of $-\kappa_\mu$
within the $2\sigma$ ranges of both $\Delta a_\mu$ and $\Delta a_e$  are much smaller than 
those within the $2\sigma$ ranges of $\Delta a_\mu$. This is because $\kappa_\tau$ is required to be large enough
to explain $\Delta a_e$, and for such large $\kappa_\tau$, $\Delta a_\mu$ favors a relative small $\kappa_\mu$.

\begin{figure*}[tb]
%\begin{center}
\includegraphics[width=8.cm]{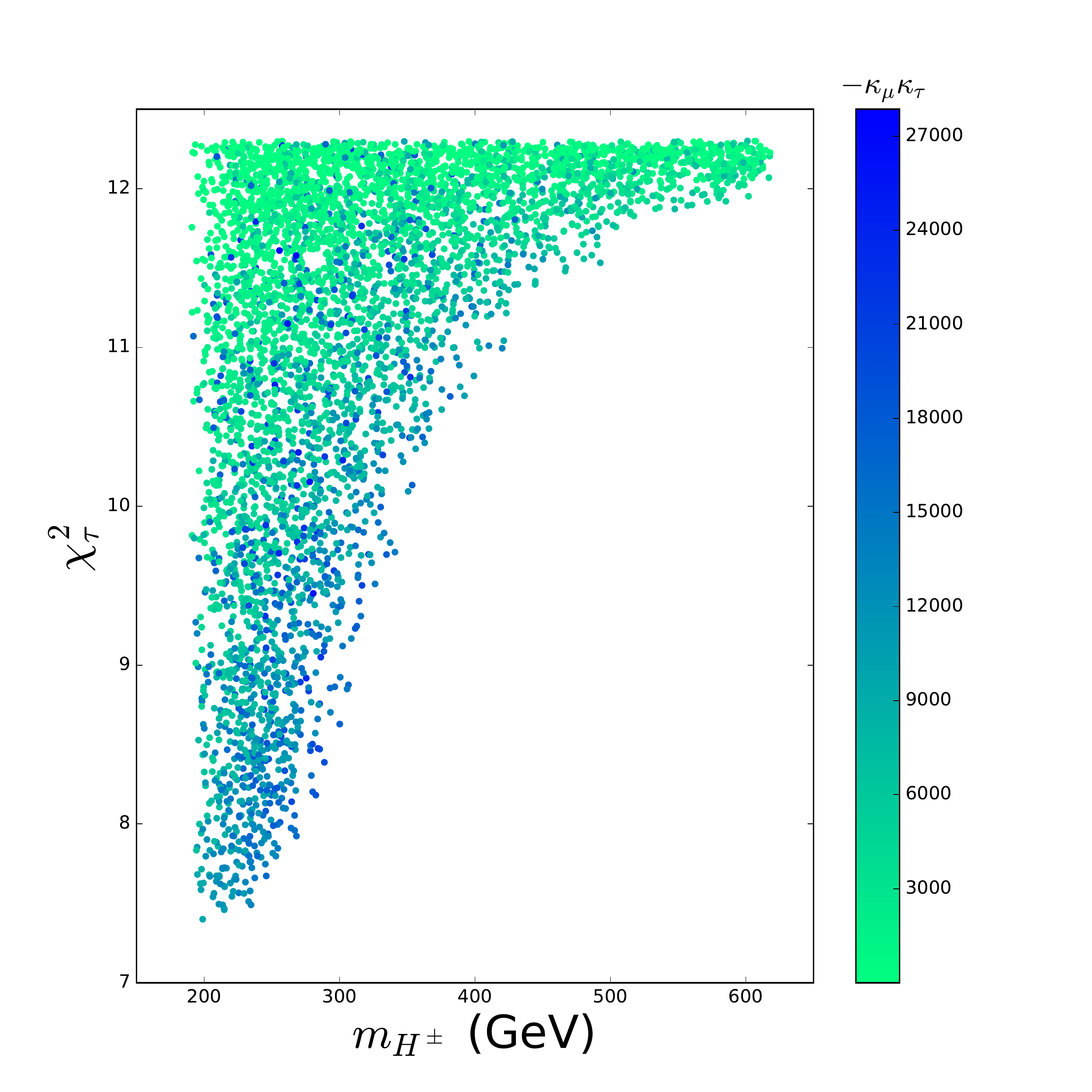}
\includegraphics[width=8.cm]{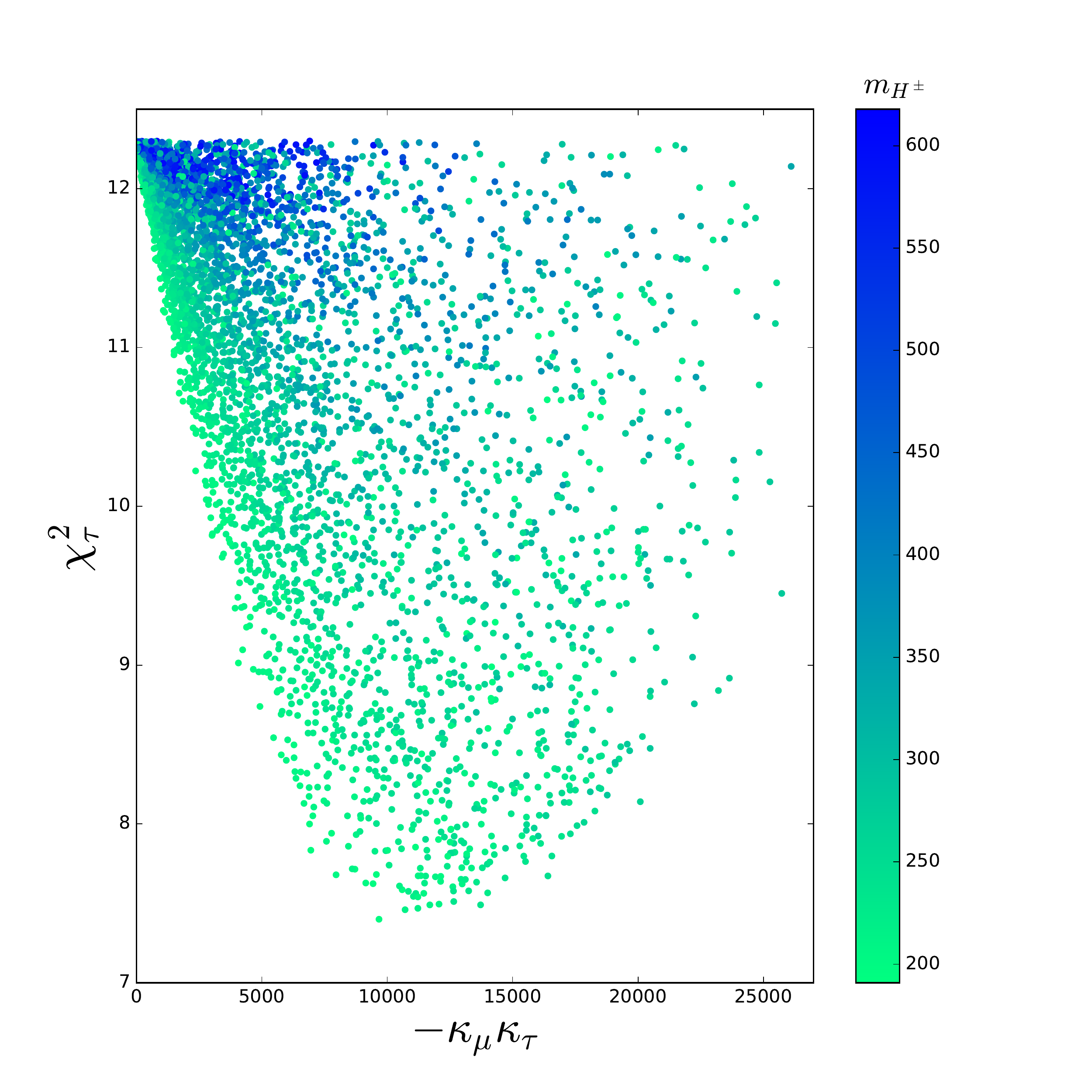}
%\end{center}
\vspace{-0.4cm}\caption{The surviving samples fit the data of LFU in $\tau$ decay with $\chi_\tau^2<$ 12.3. 
All the samples satisfy the constraints of "pre-muon $g-2$".}
\label{tauyes}
\end{figure*}

\begin{figure*}[tb]
%\begin{center}
\includegraphics[width=13.cm]{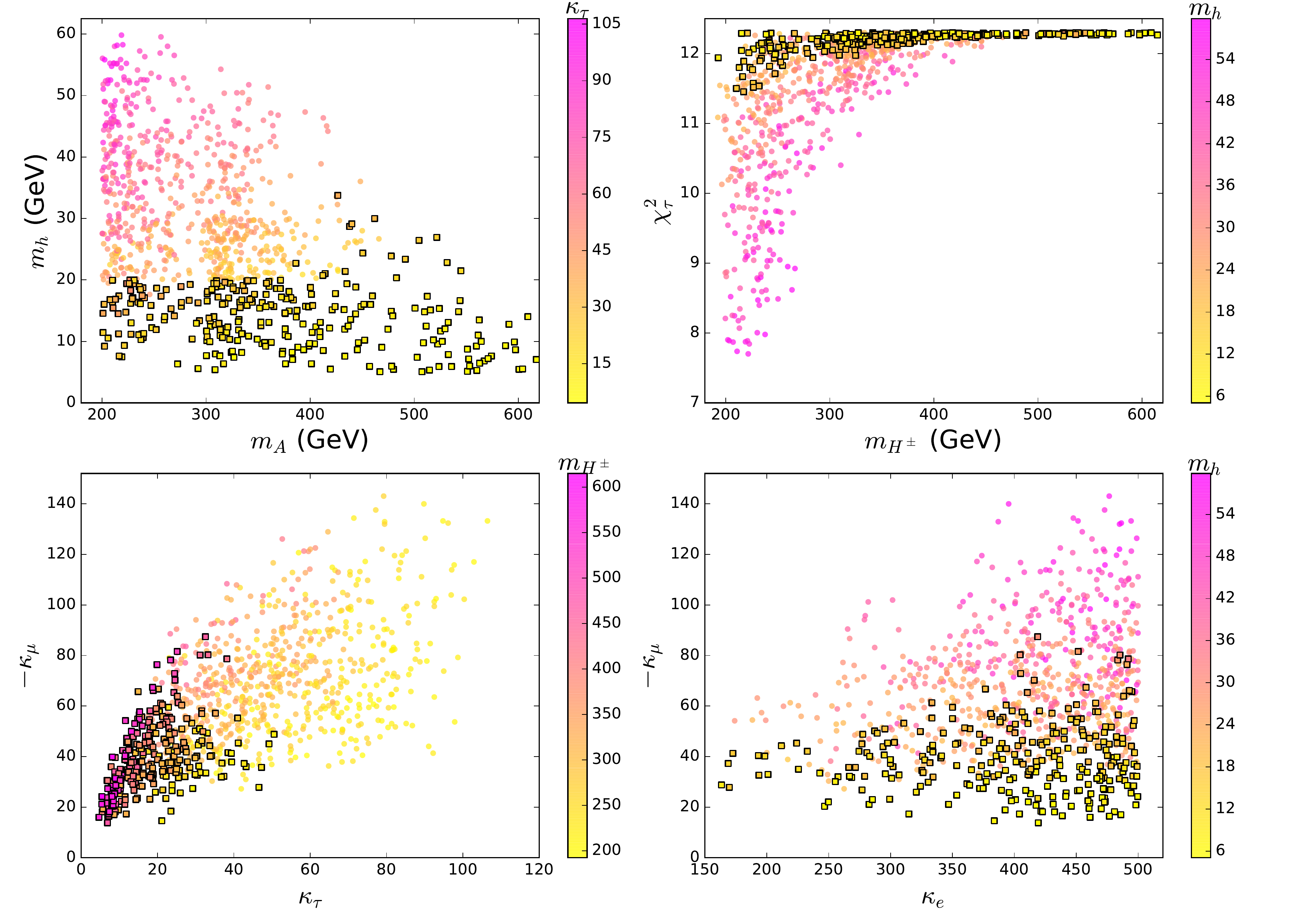}
%\end{center}
\vspace{-0.4cm}\caption{The allowed samples (squares) and excluded samples (bullets) by the direct search limits from the LHC at
95\% confidence level. All the samples satisfy the constraints of "pre-muon $g-2$", $\Delta a_\mu$, $\Delta a_e$, $\chi^2_\tau<12.3$, and $Z$ decays.}
\label{zyes}
\end{figure*}

 After imposing the constraints of "pre-muon $g-2$", we show the surviving samples with $\chi_\tau^2<$ 12.3 in Fig. \ref{tauyes}.
Because $\kappa_\mu$ is opposite in sign from $\kappa_\tau$, the second term of $\delta_{\rm tree}$ in Eq. (\ref{tree-tau}) is positive,
which gives a well fit to $g_\tau/g_e$. Such case is not realized in the lepton-specific 2HDM and aligned 2HDM.
From Fig. \ref{tauyes}, we find that $\chi_\tau^2$ increases with $m_{H^\pm}$, and obtains a relative small value for 
the moderate $-\kappa_\mu\kappa_\tau$. The $\chi_\tau^2$ can be as low as 7.4, which is much smaller than the SM value.

In Fig. \ref{zyes} we show the surviving samples after imposing the constraints of "pre-muon $g-2$", 
$\Delta a_\mu$, $\Delta a_e$, $\chi^2_\tau<12.3$, $Z$ decays, and the direct searches at LHC.
The Fig. \ref{zyes} shows that the points with relatively larger $m_{A}$/$m_{H^\pm}$ or lower $m_{h}$ can escape the direct searches. 
The production cross sections at the LHC decrease with heavier $A$/$H^\pm$, and  this region can be further detected 
with higher luminosity and collision energy. For the light $h$, the $\tau$s from $h$ in
 decays become too soft to be distinguished at detector, while the $\tau$s from $h$ in $A/H^\pm$ decays are collinear 
because of the large mass splitting between $h$ and $A/H^\pm$. Meanwhile, the $A/H^\pm \to h Z/W^\pm$ decay modes dominate 
the $A/H^\pm$ decays in the low $m_{h}$ region. Thus, in the region of $m_{h}<20$ GeV, the acceptance of above signal region for final state of 
two collinear $\tau$ + $Z/W$ boson quickly decreases.

The upper-right panel of Fig. \ref{zyes} shows that the $\chi^2_\tau$ is required to be closed to the SM value since the 
 multi-lepton event searches at the LHC favor large $m_{H^\pm}$ or small $m_{h}$.
 For small $m_{h}$, the muon and electron anomalies favor small $\kappa_\tau$ and $-\kappa_\mu$. 
As a result, the new contribution to $\chi^2_\tau$ are sizably reduced.
The  $\chi^2_\tau$ is allowed to be as low as 11.5 for $m_{H^\pm}$ around 200 GeV, which is visibly smaller than the SM value, 12.3. 
The Fig. \ref{zyes} shows that the parameters of the Yukawa couplings are favored in
the region of 15 $<-\kappa_\mu<90$, 5 $<\kappa_\tau<50$, and 160 $<\kappa_e<500$.

\section{Conclusion}
 In the lepton-specific inert two-Higgs-doublet model, we consider relevant theoretical and experimental constraints, especially for  
the multi-lepton event searches at the LHC, and discuss the possibilities of explaining the new muon $g-2$ 
anomaly reported by the Fermilab and electron $g-2$ anomaly. We find that the muon and electron $g-2$ anomalies can be
 explained simultaneously in the region of 5 GeV $<m_h<$ 35 GeV, 200 GeV $<m_A<620$ GeV, 190 GeV $<m_{H^{\pm}}<620$ GeV, 
15 $<-\kappa_\mu<90$, 5 $<\kappa_\tau<50$, and 160 $<\kappa_e<500$.

\section*{Acknowledgment}
This work was supported by the National Natural Science Foundation
of China under grant 11975013 and 11875062, and by the Project of Shandong Province Higher Educational Science and
Technology Program under Grants No. 2019KJJ007.

%%%%%%%%%%%%%%%%%%%%%%%%%%%%%%%%%%%%%%%%

\end{document}